# Stock prices assessment: proposal of a new index based on volume weighted historical prices through the use of computer modeling

Tiago Colliri[1], Fernando F. Ferreira[1]

[1]Universidade de São Paulo, Brazil
tcolliri@gmail.com, ferfff@gmail.com

**Abstract.** The importance of considering the volumes to analyze stock prices movements can be considered as a well-accepted practice in the financial area. However, when we look at the scientific production in this field, we still cannot find a unified model that includes volume and price variations for stock assessment purposes. In this paper we present a computer model that could fulfill this gap, proposing a new index to evaluate stock prices based on their historical prices and volumes traded. Besides the model can be considered mathematically very simple, it was able to improve significantly the performance of agents operating with real financial data. Based on the results obtained, and also on the very intuitive logic of our model, we believe that the index proposed here can be very useful to help investors on the activity of determining ideal price ranges for buying and selling stocks in the financial market.

## 1. Introduction

The correlation between volume and stock price change was first suggested in 1959, by Osborne [22], and later developed by Clark [4], Tauchen and Pitts [25], and Harris [12], who presented evidences for this correlation by analyzing data series from the New York Stock Exchange as well as from cotton and treasure bill future contracts. In 1964, Godfrey, Granger and Morgenstein [10] also presented new findings about the existence of a positive correlation between daily volume and daily price variation. And empirical studies made by Ying [26] and Crouch [6] with stock market aggregates also contributed to support the correlation between the two. Since then, many other works contributed to reinforce this concept, as appointed by Karpoff [19] in his survey. However, in spite of many studies that have been made in this area, we still do not have a unified model that aggregates traded volume and price variations for stock prices assessment purposes, especially for long-term investment horizon.

Empirically speaking, we could enunciate that the ideal price for buying a stock in the market would be the smallest possible among the prices paid for those who already bought that stock in the past. In other words, if we could put all the prices paid for the respective buyers that currently still have positions in that stock in an x axis, so the current price of the stock would be more attractive as more it is on the left side of the x axis, when compared with the other prices. Continuing with this thought process, let us now suppose that on the y axis we could have all the volumes currently owned by the investors, each one of them measured as a proportion of the total number of shares available for that stock in the market and whose total sum is equal to 1. In such a way that, looking at the graphic, we could know all the prices paid by the current investors at the time they bought the stock and also how much of the total volume of shares is currently owned by them. If we could have such of graphic, then to determine if the current price of a stock is attractive or not, we would just need to see which proportion of the total volume of the stock was traded on the left side of the current price in the x axis. As more on the left side of the prices paid by the current buyers the current price is, so more attractive it would be for the prospective investor. And, on the other hand, as more volumes there would be on the left side of the current price on the x axis and, consequently, more on the right side of the x axis the current price is, so less attractive this stock would be for the prospective investor. Following this same logic, a 'neutral' price would be considered as the price in which we have one half of the total volume of the stock traded below that price (on the left side of the x axis) and the other half of the total volume of the stock traded above that price (on the right side of the x axis). The main objective of the present model is exactly to provide such of graphic. And, to do so, we simply 'reconstruct' all the trading history of the stock, since its IPO (initial public offering) until present. At first, it may sounds like a very complex task but, actually, it is not, and can be easily made with a computer model, as will be showed here.

Regarding to the organization of this paper, in section 1 we have this brief introduction. In section 2 we describe our index with more details, introducing the computer model that was used for its generation, and also the methodology we used for testing it, which was the multiagent simulation (MAS). In section 3 we show and discuss the results obtained by applying our model to some real financial data from the Brazilian stock market and testing it through the agents that were generated. And, in the section 4, we have the final conclusions.

## 2. Methodology

To describe our model we start by introducing its input data and output data. Then we show the calculation methods behind the generation of the index. And, at the end, which technique we used for testing the index accuracy is provided.

### 2.1 Input and output data

The concept of volume used in this model is the one known as 'free float', which represents the quantity of shares effectively available for trading in the stock market. Thus, the total volume of shares of a company was subtracted by the quantity of shares owned by the majority shareholders, which are not traded on a day-to-day basis in the stock market, according to the shareholding structure of each company. In the case of changes in the shareholding structure along the time, then the free float is also altered. For example, if the company makes a second IPO, then the number of available shares in the stock market (free float) will be refreshed at that date, by been added with the quantity of new shares emitted by the company. And on the other hand, if the company buys some quantity of its own shares in the market and, later, cancels those shares, then the free float will be refreshed this time by subtracting that quantity of shares, at the same date of its cancellation.

Regarding to the input data required to run the model, it requires the following:

- the historical average daily prices, which can be obtained by dividing the total daily traded volume measured in currency units by the total daily traded volume measured in quantity of shares (more accurate) or by taking the average between the highest and the lowest daily prices (less accurate),
- the historical daily traded volume, and
- possible operations that affects the free float, such as new IPO's and shares cancellations made by the company during the period.

Once that we have this input data, we then run the model and it will returns the historical current volume weighted average price (VWAP), which represents what would be the 'neutral' price for buying that stock, for each day of the period. And it also provides a graphic in which, in the x axis, we have the prices paid in the past by the current investors and, in the y axis, we have the respective remaining volumes owned by these current investors, according with the price they paid, measured as a proportion of the free float and whose total sum is equal to 1.

### 2.2 Calculation methods

Before running the model, we organized the input data as following: the volumes and prices are sorted in ascending order according with their chronological dates, and separated into two different arrays, one array $v$ for the volumes and another array $p$ for the prices. Then we introduced, at the first row of the array $v$, the number that represents the total number of shares emitted by the company in its first IPO and, on the row of the array $p$, we introduced the respective price of the shares at the moment of the IPO. If there were any operations, since the IPO until the last day verified in the data, that altered the free float of the company, such as new IPOs or canceling of shares, then we also introduced these data into the arrays, always respecting the chronological order. For example, if the company canceled 1,000 shares (hypothetically) 30 trading days after its IPO, so we introduced the number 1,000 at the row 31 of the array $v$ (it would be 31 because the first row corresponds to the IPO and so it does not counts as a trading day). And we also introduced its respective price at the same row of the array $p$. In this model we considered the current price of the stock as the price for cases of cancelations of shares. When the computer model will read this line it will know it is a cancelation of shares (by adding some specific signal on it, as a signal of minus before the number, for instance) and then will subtract this number 1,000 from the current free float of the stock. Once that we are sure that we introduced all the possible alterations occurred on the free float during the period, we then run the model.

The computer model starts by reading the first rows of the arrays $v$ and $p$, which brings the free float generated by the first IPO of the company $vt$ (which is also the first volume $v_0$) and the respective initial price $p_0$ of the offer. We then record these two numbers $v_0$ and $p_0$ in two different arrays $vn$ and $pn$, respectively. That would means that, making a link with the real market, at this time we have one hundred percent of the investors bought at the same price. Then, on the second rows of the arrays $v$ and $p$, it will read the total volume $v_1$ and the average price $p_1$ traded at the first day of negotiations in the stock market, and it will record it as follows:

$$vn = [v_0, \ v_1] \quad \text{and} \quad pn = [p_0, \ p_1] \ .$$

Now, again making a link with the real market, we know that $v_1$ shares of the stock were sold by the average price $p_1$. And we also know that, because we are still on the first day of negotiations, those who sold these $v_1$ shares bought them at the day before, paying $p_0$ for them. So now, before reading the next rows of the arrays, the model subtracts $v_1$ from $v_0$, in order to keep the total sum of the array $vn$ equal to $vt$ (which is the free float). At this point, it is worth to remember that we are dealing with the daily average prices traded here, and that is the reason we can assume that everyone bought the stock for the same price at that day. Now, from the third negotiation day on, comes the uncertainty we had to deal with in order to build our model: from this day $t$ on, we know that $v_i$ shares were sold for the average price $p_t$, but what we do not know is how much were paid for these same $v_i$ shares at the moment they were bought, in the past. Or, thinking on a real market basis: which ones from all the current investors is selling these shares and leaving the stock in order to allow these new investors to buy their shares? This is a very crucial question for the model because in its answer is based the calculations that will be made to refresh the values of the pre-existing volumes in $vn$. To solve this problem we considered that, after each new day of trading, each value $v_i$ in the array $vn$ will be refreshed as follows:

$$v_{i_{t+1}} = \ v_{i_t}.(1 - \tfrac{v_{new}}{vt}) \ ,$$

where $v_{new}$ is the new value added to the array and refers to the total volume traded at the day $t+1$. Thus, the past traded volumes are subtracted proportionally to their respective amounts. And, as the model continues to run, some of these volumes will 'disappear' (keep diminishing until zero) at the same time that the new ones (which represent the new trading days) will be placed in different locations of the x axis (because they were traded at different prices). And this dynamic process continues until the last line of the input data is read, which represents the last trading day of the simulation. We would like to emphasize the fact that the use of this formulae also allows the model to maintain the total sum of the array $vn$ always equal to $vt$ (the free float), which we consider as a very important issue on the intuitive basis of our model. This same importance we give to the possible alterations that can occur with the value of $vt$ along the period, such as new IPOs or canceling of shares. The model has to deal specifically with each one of these cases, maintaining the value of the free float $vt$ as close as possible of its real value, based on the real market. That means that, for example, if the company makes a new IPO at a certain date, then when the model reads this information (which is inserted in the input arrays $v$ and $p$, respecting the position of its chronological date) it must add this new amount of shares that were emitted to the value of $vt$, besides also adding the price and the volume (the same new amount of shares) of the IPO in the arrays $pn$ and $vn$, respectively.

The volume weighted average price (VWAP) represents what would be the current 'neutral' price of the stock, in which we have one half of the total volume (free float) bought below that price and the other half of the total volume bought above that price; and it is calculated by:

$$VWAP = \sum_{i=1}^{n} p_i . v_i \ , \quad \text{where } n \text{ is the total number of current remaining volumes } v_i \text{ in the array } vn.$$

And, finally, our index is obtained by:

$$\rho = (VDI - VDS) / vt \ , \quad \text{where}$$
$VDI$ = the sum of all current remaining volumes $v_i$ which price is below the current price and
$VDS$ = the sum of all current remaining volumes $v_i$ which price is above the current price.

Consequently, the values of the index $\rho$ will vary from -1 until +1, where a value of -1 means that all the investors (one hundred percent of the past traded and still remaining volume) are currently having losses (because the current price of the stock is below the price they paid) and a value of +1 means that all the investors (one hundred percent of the past traded and still remaining volume) are currently having profits (because the current price of the stock is above the price they paid).

At the end we will have, according to our model, the expected current volume weighted average price (VWAP) of the stock, as well as a graphic with all the still remaining traded volumes from the past, in the y axis, with the respective prices at each of them were traded, in the x axis. And we will also have the evolution of our index, which represents the relation between the current percentage of the total volume (free float) that are having losses and the current percentage of the total volume that are having profits, along the time.

### 2.3 Testing the model: the use of multiagent simulation (MAS)

As long as we know that, of course, the final data provided by our model cannot be considered as 'real data', and since it is also not possible to obtain such of data in the real market to make a comparison with our data (otherwise there would not be any reason for developing such of model), we then used the technique of multiagent simulation (MAS) for testing the credibility of the model.

We created 39 agents to operate with real financial data from Bovespa – the Sao Paulo Stock Exchange. All the agents would buy and sell a stock according exclusively with our index behavior. Each agent received a parameter, which varies from 0.025 until 0.975, and he would buy a stock only when its index value reaches the value of his parameter on the negative side. And, alternately, he would sell a stock only when its index value reaches the value of his parameter on the positive side. Thus, the first agent, whose parameter is 0.025, would buy a stock every time the index value is equal or below -0.025 and he would sell it every time the index value is equal or above +0.025. Hence, the first agent will make much more operations than the last agent, whose parameter is 0.975.

We believe that, since our model is intended for very practical purposes of financial investment, it is useful to stress what these parameters would mean in the real market. As we know, the index varies from -1 until +1, where a value of -1 would means that all volumes from the free float of the stock are currently having losses, because the current price is below the price paid by the current investors when they bought their shares, in the past. And, on the other extreme, when our index reaches the value of +1, it would means that all volumes from the free float are currently having profits, because the current price is above the price the current investors paid for their shares, when they bought the stock in the past. This estimation, of how much the current investors paid for their shares in the past, is possible due to the method of calculation used in the model for refreshing the past volumes at each new trading day, as we saw in section 2.2. So, if an agent buys the stock only when the index reaches the value of -0.5, for instance, that would means that, at that exactly time, the model estimates that the sum of volumes that are currently having losses exceeds the sum of volumes that are currently having profits in one half of the total volume (or free float). As we already know, the current price of the stock will be more attractive as lower the value of the index is, where the most attractive price would be when the index is -1, and the most unattractive price would be when the index is +1. In this sense, it is expected that the last agents, whose parameters are bigger, besides making a smaller number of operations during the period, will also have longer and more profitable operations, when compared with the first agents, whose parameters are smaller and, hence, it is expected that, besides making a bigger number of operations during the period, their operations will be shorter and less profitable. Making a parallel with the real market, we could say that the characteristics of the first agents would be more similar to those from short-term investors, while the last agents would, here, represent the long-term investors.

The returns of all the operations made by each agent is recorded and accumulated, in such a way that, after the simulation, we can have a list of all their operations, with their respective buying and selling prices, as well as their accumulated returns during the period. At the end of the simulation, we took these returns and made a comparison among them, to measure the efficiency of each value of the parameter in terms of return. And, as our main objective, we also compared the returns obtained by our agents with the overall return offered by the stock during the same period.

## 3. Results and discussion

We applied our model to 3 different stocks from the Sao Paulo Stock Exchange (Bovespa), which were the following:

- JBSS3 (weight 0.669% on Ibovespa index). Industrial sector: food. Period: from April 2, 2007 until August 21, 2012.
- MRFG3 (weight 0.895% on Ibovespa index). Industrial sector: food. Period: from June 29, 2007 until August 21, 2012.
- OGXP3 (weight 2.583% on Ibovespa index). Industrial sector: oil and gas extraction. Period: from April 4, 2007 until August 21, 2012.

On figures 1 to 5 we show the results obtained for JBSS3.

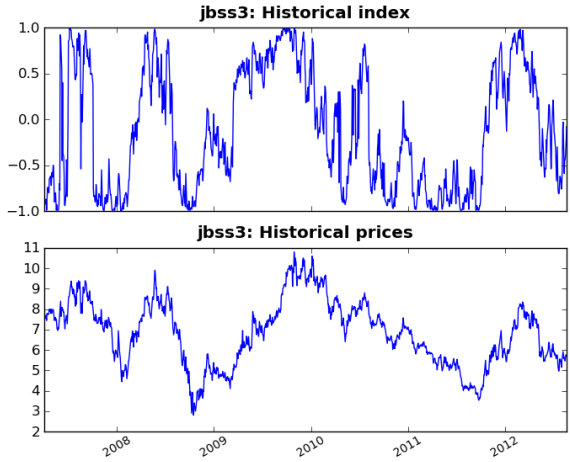

**Fig. 1.** JBSS3: Historical index and historical prices

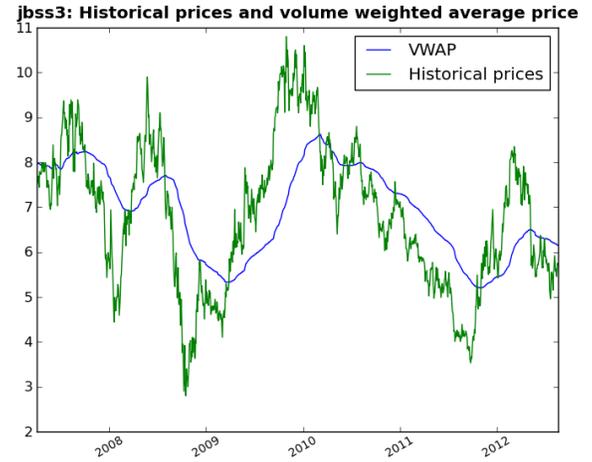

**Fig. 2.** JBSS3: Historical prices and volume weighted average price (VWAP)

In figure 1 we can see that there is a synchronicity between the behaviors of the price and our index. And that the index was capable of emphasize the highs and lows of the prices, which is exactly the most useful characteristic about it from an investor's point of view, because it helps to determine price levels for buying and selling a stock. Just remembering here that as more close the index value is from -1, more the current investors are having losses, according with the model estimations. And a value of -1 would mean that one hundred percent of the current investors are experiencing temporary losses, which happens either when the stock price is in an historical low, or when the investors who bought the stock for a lowest price in the past already sold their shares and, hence, have left the stock and do not compose the current free float. The same occurs in the other extreme of the index: when its value is very close to +1 it means that almost all investors are currently having profits, according with our model estimation, and so this probably would not be a good moment for buying that stock.

In figure 2 it is possible to see the behavior of the volume weighted average price (VWAP) along the time, compared with the historical quotes of the stock. It is interesting to note that, besides the significantly difference between the calculations, its behavior resembles other already existing averages used for technical analysis in the market, such as the simple moving average and the exponential moving average.

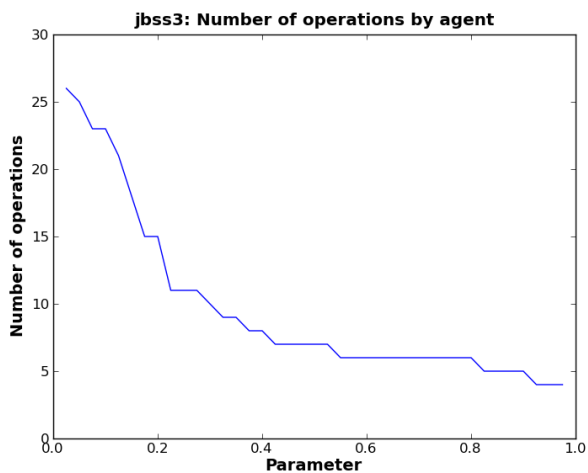

**Fig. 3.** JBSS3: Number of operations by agents operating exclusively with our index, based on their respective parameters

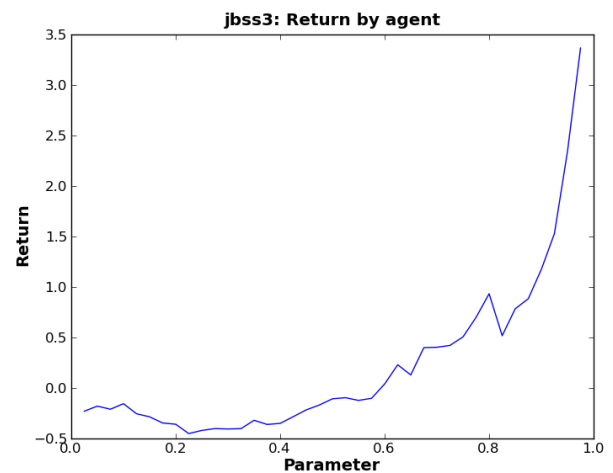

**Fig. 4.** JBSS3: Total return by agents operating exclusively with our index, based on their respective parameters

In figure 3 are shown the total number of operations made by our agents while operating exclusively based on our index behavior. One operation here is considered as a buying and a selling order. As it was expected, the first agents, whose parameters are smaller than those from the last ones, made much more operations during the period.

In figure 4 we can see the total return obtained by each agent, and what stands out here is the total return of our last agent, which was of 337% during the period. And if we consider that the overall return offered by that stock during the same period was of -28%, so it means that our index was capable of increase significantly the performance of our last agent. We also underline that this same agent made only 4 operations during the period, which also contributes to show the high optimization reached by his performance. Looking at the agents operations list, we saw that the operations made by the last agent were: R$ 7.42 (buy), R$ 8.64 (sell), R$ 6.95 (buy), R$ 9.56 (sell), R$ 4.61 (buy), R$ 9.02 (sell), R$ 6.00 (buy) and R$ 8.36 (sell). It is also worth noting in this graphic the tendency of, as the parameters used by the agents for operating in the market get higher, their returns get higher as well. We believe that this feature helps to support the empirical basis of our model, since the last agents operate according with the extreme values of our index.

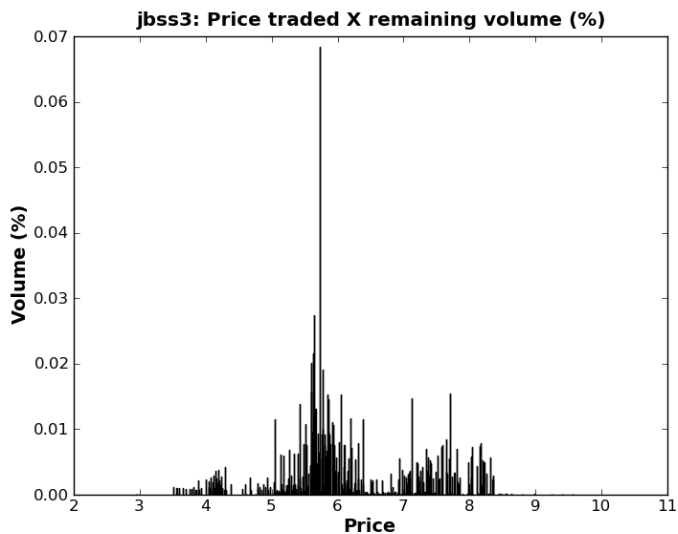

**Fig. 5.** JBSS3: Current remaining volumes (%) and their respective traded prices

In figure 5 there is a representation of what would be the final remaining volumes traded in the past with their respective prices, measured as a proportion of the current free float. If, by one hand, we know that this representation cannot be considered an exactly representation of what would be its real data (if we could have access to it to compare both), by the other hand, we claim that, given the very intuitive logic behind the model's simple calculation methods, the objective here is to have a model capable to provide what could be considered as a rough approximation of the reality. And, in this sense, we believe that this representation could be very useful for investors, especially when analyzed together with the volume weighted average price (VWAP), which, here, was equal to R$ 6.14 at the end. The logic behind this analysis is very simple: the current price of the stock would be more attractive as more it is on the left side of the x axis and (the most important) lesser and smaller columns there would be at its left; and where the VWAP would represent a 'neutral' price, with fifty percent of the remaining traded volumes at its left and the other fifty percent at its right. Looking at the histogram we can see that there is a price, of R$ 5.73, which stands out from the other prices, been responsible, alone, for 6.83% of the composition of the free float. Looking at the historical data of this stock, we could see that the explanation for this was in the fact that on May 31, 2012 there was a very high level of liquidity for that stock, when its traded volume reached almost 10 times the average traded volume for the period.

On figures 6 to 10 there are the results for the stock MRFG3.

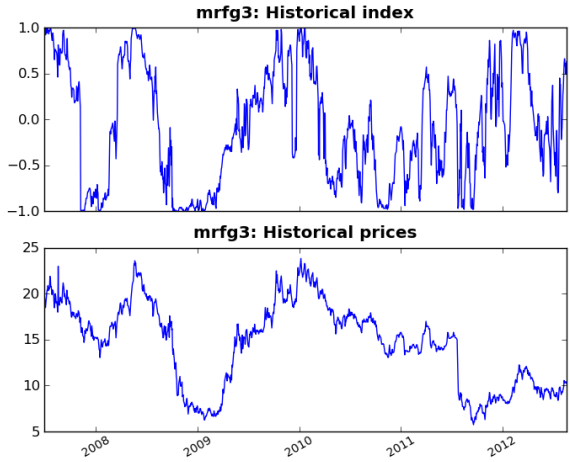

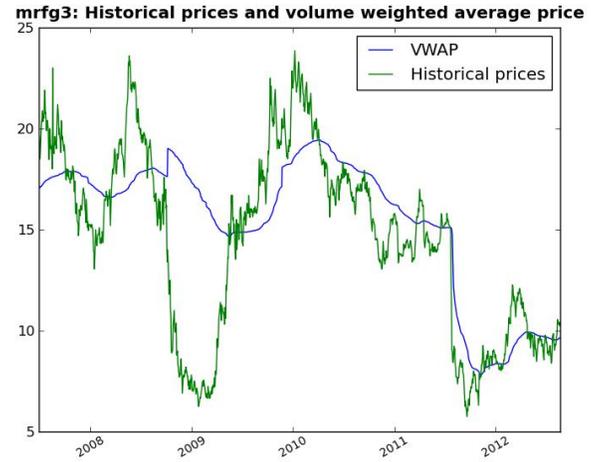

**Fig. 6.** MRFG3: Historical index and historical prices

**Fig. 7.** MRFG3: Historical prices and volume weighted average price (VWAP)

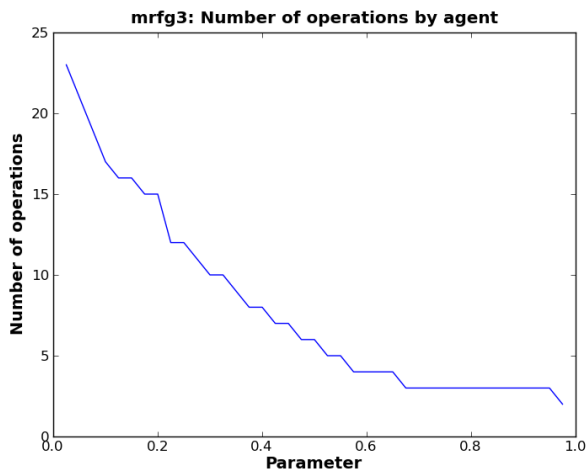

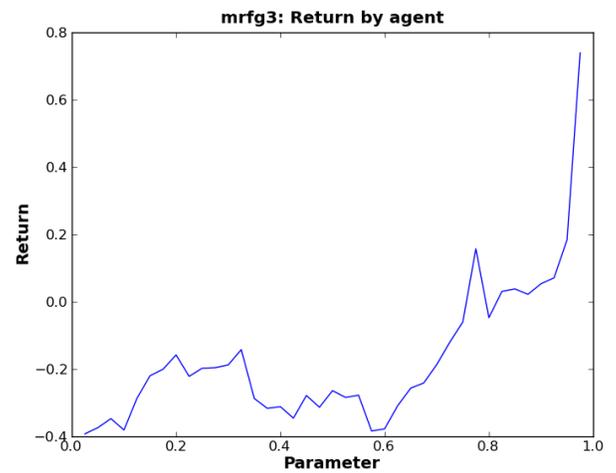

**Fig. 8.** MRFG3: Number of operations by agents operating according exclusively with our index, based on their respective parameters

**Fig. 9.** MRFG3: Total return by agents operating according exclusively with our index, based on their respective parameters

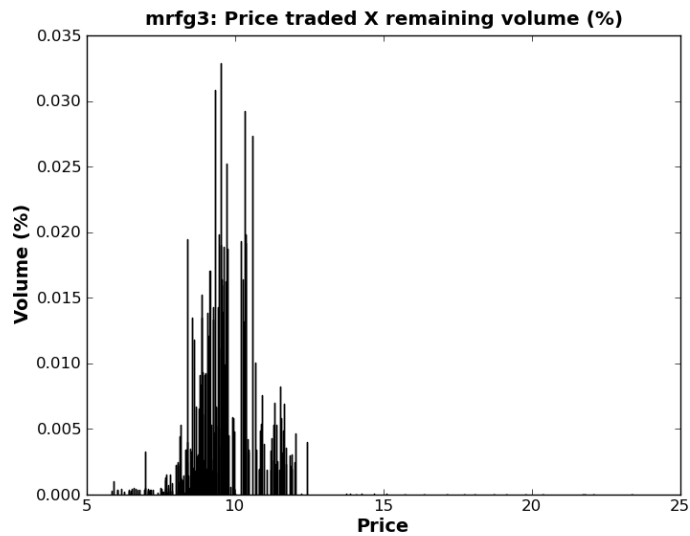

**Fig. 10.** MRFG3: Current remaining volumes (%) and their respective traded prices

From this series of figures, we observe that the main features occurred with the stock JBSS3 still remained here (e.g., the synchronicity between the stock price and the index behaviors and the major gains obtained by our last agents). This stock had a very abnormal movement between July 27, 2011 and August 8, 2011, when the stock price fell down by 45% in only 10 trading days, and this made our index also drop heavily on this period, as we can see in figure 6. But, besides this, we can say that the index still worked pretty well, since our last agent, with only 3 operations, obtained 74% of total return, while the stock price, during the same period, fell down by 39%. In this case, the figure 8 appoints 2 operations for the last agent, and it happened because his last operation is still 'open' (he didn't sell his last shares before the end of the simulation). Checking on the operations list, we could see that his operations with this stock were the following: R\$ 16.65 (buy), R\$ 21.1 (sell), R\$ 12.67 (buy), R\$ 22.27 (sell) and R\$ 13.26 (buy), where this last buying operation occurred on the day the stock price fell down abnormally.

Now, on figures 11 to 15, we show the results obtained for the stock OGXP3.

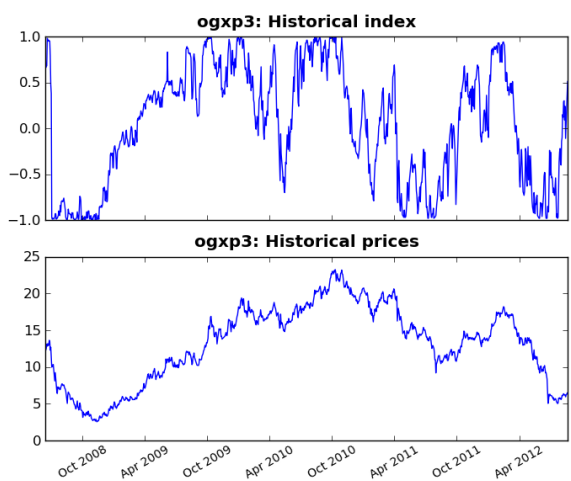

**Fig. 11.** OGXP3: Historical index and historical prices

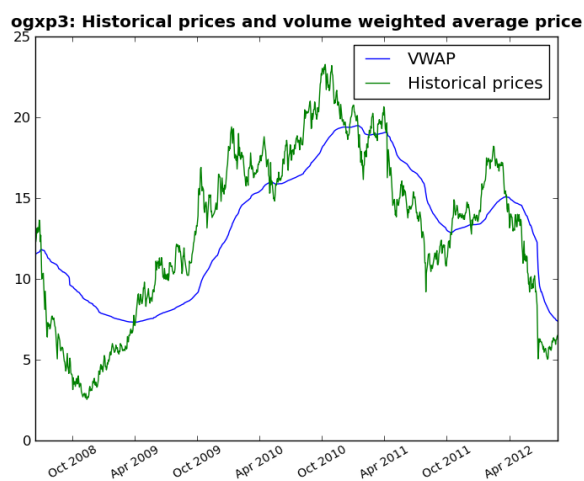

**Fig. 12.** OGXP3: Historical prices and volume weighted average price (VWAP)

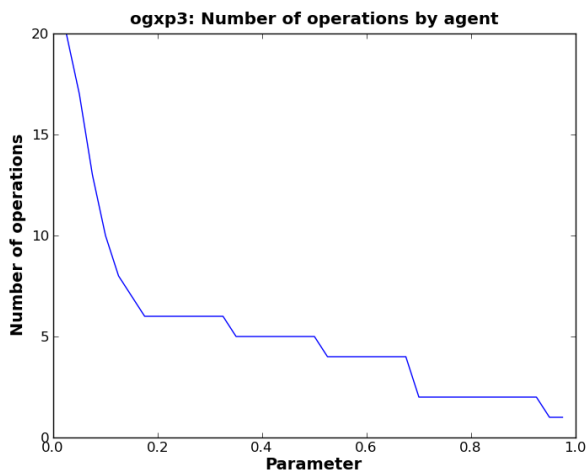

**Fig. 13.** OGXP3: Number of operations by agents operating according exclusively with our index, based on their respective parameters

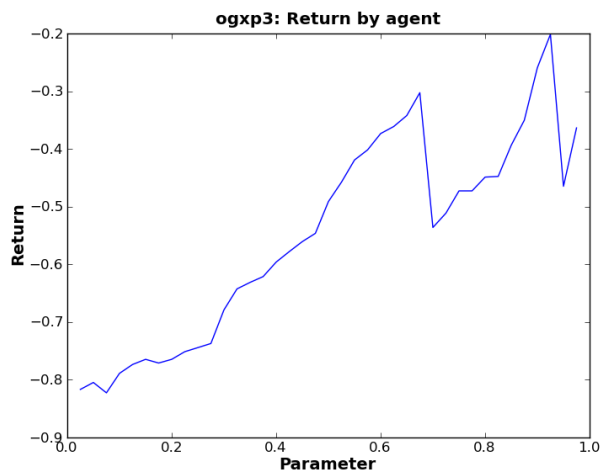

**Fig. 14.** . OGXP3: Total return by agents operating according exclusively with our index, based on their respective parameters

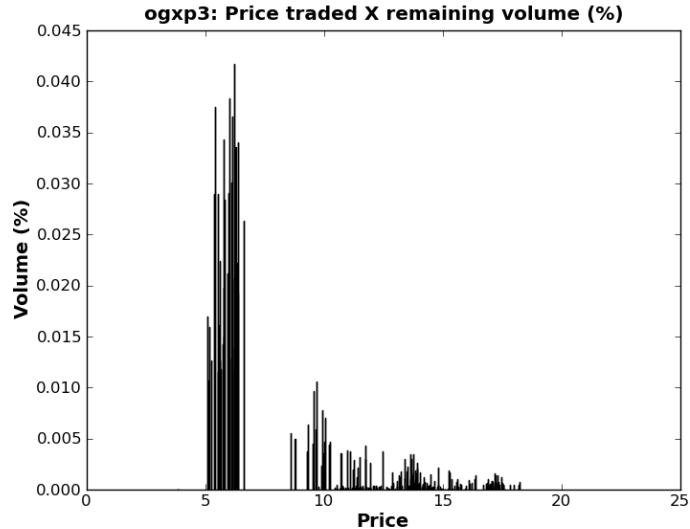

**Fig. 15.** OGXP3: Current remaining volumes (%) and their respective traded prices

For the stock OGXP3 the last agent presented a double digit negative return, which was mainly caused by an also abnormal movement of this stock price, occurred on June 27 and 28, 2012, when the price felt over 45 percent in only two days. Recent drastic movements in the stock price can affect significantly the performance of our agents, since when the simulation is over, if an agent has an open position (is bought and did not sell his shares yet) in any stock, so his position is then forcibly closed by the program in order to calculating his accumulated return during the period. It is also possible to verify that the main features of the graphics observed in the other stocks still remained here.

On the table 1 we show the overall results obtained in our tests. As we can note, for all the stocks used here the last agents of our simulations reached higher returns than the returns offered by the movement of the prices of the stocks during the period.

**Table 1**. Stocks used in the tests and the respective returns obtained by the last agents

| Stock | Coefficient of Variation of Daily Prices | Overall Price Return (%) | Return of the Last Agent (%) |
|---|---|---|---|
| JBSS3 | 0.23 | -28 | 337 |
| MRFG3 | 0.31 | -39 | 74 |
| OGXP3 | 0.40 | -42 | -36 |
| **Average** | | **-36.2** | **125** |

We chose to use the returns obtained by the last agents for representing the performance provided by the use of the index for the reason that they operate with the real market data according exclusively with the most extreme values of our index, and hence we believe that if they have succeeded in their operations then it can be seen as a very good sign that the logic behind the calculation of the index has empirical support.

## 4. Summary and conclusions

In this paper we proposed a new index for evaluating stock prices in the financial market that takes in account historical traded volume data. The calculus used for generating the index can be considered relatively simple, and only needs the help of a computer model for its implementation. Through the technique of agent based simulation, applied to the real financial data utilized here, it was possible to establish two main conclusions: (1) the use of the index was capable of increasing significantly the performance of the agents, in terms of their returns, and (2) the index seems to work better, i.e., increase the optimization of the operations, when agents operated in the market according with its extremes, which are the values nearer -1 and +1.

Our idea, by proposing this index, is that it can be used as a tool for the technical analysis of financial stocks. As it is known in the financial market, the technical analysis takes in account mainly the behavior of the stock price for evaluating whether the current price of a stock is attractive or not. We consider that the main contribution of our model is the inclusion of the past traded volumes in this analysis, in such a way that it takes in account the already accepted correlation existent between volumes and price variations and unifies both of them for practical assessment purposes. We believe that this index can be very useful for investors on the activity of determining ideal price ranges for buying and selling stocks in the financial market, especially when used together with fundamental analysis, which takes in account the real financial data of a company, such as its current cash flow and forecasted profits.

Although the results obtained here can be considered as a very positive sign about the index efficiency, we would like to suggest, as a recommendation for future works, testing the behavior of the index with a larger number of different stocks. Given the very simple and intuitive logic of the model, it is expected that the synchronicity between the index and the stock price behaviors and, consequently, the financial performance optimization verified here will continue to remain when applied to different stocks data. It would also be interesting considering different forms of estimating de distribution of the selling orders among the current investors, besides the volumes proportional distribution that was used here, which is made at the step in the calculations where the pre-existing traded volumes are subtracted in order to allow the new traded volumes to compose the free float.